\documentclass[12pt]{article}
\usepackage{graphicx,amssymb,amsmath,empheq}
\usepackage{amsthm}
\usepackage{empheq}
\usepackage{subfig} 

\usepackage{hyperref}
\hypersetup{colorlinks = true, linkcolor=black, citecolor=black, urlcolor=blue}
\usepackage{url}

\theoremstyle{plain}

\usepackage{tikz}
\usepackage{tikz-cd}
\usetikzlibrary{arrows}
\usetikzlibrary{intersections}
\usetikzlibrary{shapes.geometric}
\usetikzlibrary{decorations.pathmorphing, patterns,shapes}

\usepackage[nosort]{cite}

\newenvironment{defn}[1][Definition]{\begin{trivlist}
\item[\hskip \labelsep {\bfseries #1}]}{\end{trivlist}}

\renewcommand{\bar}{\overline}
\renewcommand{\tilde}{\widetilde}

\renewcommand{\leq}{\leqslant}
\renewcommand{\geq}{\geqslant}
\renewcommand{\Re}{\operatorname{Re}}
\renewcommand{\Im}{\operatorname{Im}}

\newcommand{\Tr}{\operatorname{Tr}}
\newcommand{\nn}{\nonumber}

\newcommand{\VF}{\Upsilon}
\newcommand{\OTOC}{\operatorname{OTOC}}
\newcommand{\BOX}{\operatorname{BOX}}
\newcommand{\dkap}{\delta\kern-1.25pt\varkappa}

\newcommand{\R}{{\mathrm{R}}}
\newcommand{\A}{{\mathrm{A}}}
\newcommand{\W}{{\mathrm{W}}}

\newcommand{\SL}{\operatorname{SL}}

\newcommand{\T}{\operatorname{\textbf{T}}}

\newcommand{\RR}{\mathbb{R}}

\newcommand{\kap}{\varkappa}

\newcommand{\calF}{\mathcal{F}}

\newcommand{\calH}{\mathcal{H}}

\newcommand{\calJ}{\mathcal{J}}

\newcommand{\calO}{\mathcal{O}}

\newcommand{\Sch}{\operatorname{Sch}}
\newcommand{\sgn}{\operatorname{sgn}}

\title{On the relation between the magnitude \\
and exponent of OTOCs}

\author{
Yingfei Gu\footnote{yingfei\_gu@g.harvard.edu}\\
\normalsize\it
Harvard University, Cambridge, MA 02138, U.S.A\\
[5pt]
Alexei Kitaev\footnote{kitaev@caltech.edu}\\
\normalsize\it California Institute of Technology, Pasadena, CA 91125, U.S.A.
\vspace{0.5cm}
}
\date{November 30, 2018}

\begin{document}

\maketitle

\begin{abstract}
We derive an identity relating the growth exponent of early-time OTOCs, the pre-exponential factor, and  a third number called ``branching time''. The latter is defined within the dynamical mean-field framework, namely, in terms of the retarded kernel. This identity can be used to calculate stringy effects in the SYK and similar models; we also explicitly define ``strings'' in this context. As another application, we consider an SYK chain. If the coupling strength $\beta J$ is above a certain threshold and nonlinear (in the magnitude of OTOCs) effects are ignored, the exponent in the butterfly wavefront is exactly $2\pi/\beta$.
\end{abstract}

\tableofcontents

\section{Introduction}

Out-of-time-order correlators (OTOCs) are a diagnostic of chaos in quantum systems~\cite{LaOv69}. This paper is concerned with the early time regime, which exists in various models with a large parameter $N$. For systems with all-to-all interactions, $N$ is simply the number of elementary parts such as spins or fermionic sites. However, a large parameter can be defined in many cases where the interaction is local. For example, in a Fermi liquid, $N$ is the number of quasiparticles in a region whose size is determined by the (inelastic) mean free path. A common behavior is that at some time scale, correlation functions obey an approximate clustering property such that all connected correlators are suppressed by a large factor $C$ proportional to $N$. At slightly larger times, the connected OTOCs grow as $C^{-1}e^{\kap t}$ until saturating at a value of the order of~$1$. It is this initial growth that we focus on.

The Lyapunov exponent $\kap$ (called so by analogy with classical dynamical systems) is very important in characterizing chaos. Under the stated conditions, it satisfies the tight upper bound $\kap\leq 2\pi/\beta$~\cite{MSS15}. The maximum value is realized for field correlators in the vicinity of a black hole, though a small negative correction occurs due to stringy effects~\cite{ShSt14}. The SYK model~\cite{SaYe93,Kit.KITP.1,Kit.KITP.2,KS17-soft} at low temperature was the first concrete example of a Hamiltonian that saturates the bound. More exactly, the OTOCs of Majorana fermions for the SYK model are characterized by these parameters~\cite{MS16-remarks}:
\begin{equation}
C\sim\frac{N}{\beta J},\qquad 1-\frac{\kap\beta}{2\pi}\sim\frac{1}{\beta J}.
\end{equation}
Therefore, the following quantity, which determines commutator OTOCs~\cite{KS17-soft}, has a finite limit at zero temperature:
\begin{equation}
r=\frac{\cos(\kap\beta/4)}{C}.
\end{equation}

We will derive an expression for $r$ that is applicable whenever the dynamic mean field approximation works. (Technically, we rely on the representation of four-point functions by ladder diagrams.) This is useful because $C$ and $\kap$ are usually computed by different methods. In principle, both parameters can be obtained by analytically continuing the four-point correlator from imaginary to real times. This technique is efficient for the SYK model at low temperature if one uses the Schwarzian approximation~\cite{Kit.KITP.2,MS16-remarks,KS17-soft} but the correction to the Lyapunov exponent is lost; a more accurate calculation of the four-point function requires more work~\cite{MS16-remarks,KS17-soft}. On the other hand, $\kap$ can be found by solving a kinetic equation involving a retarded kernel~\cite{Kit.KITP.1,MS16-remarks,MSW17}. Knowing the number $r$ allows one to use either method without losing any information. The fact that $r$ does not diverge or vanish as a function of various parameters (e.g.\ the momentum in the case of the SYK chain) has interesting physical consequences, which we will also discuss.

\section{Preliminaries}

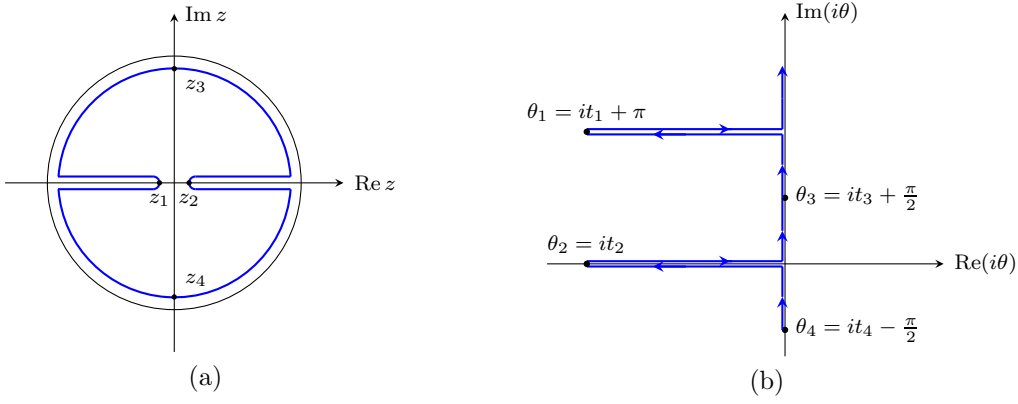
\begin{figure}
[t]
\center
\subfloat[]{
\begin{tikzpicture}[scale=0.8, baseline={(current bounding box.center)}]
\draw[->,>=stealth] (-80pt,0pt) -- (80pt,0pt) node[right] {\scriptsize $\Re z$};
\draw[->,>=stealth] (0pt,-80pt) -- (0pt,80pt) node[right]{\scriptsize $\Im z$};
\draw (0pt,0pt) circle (60pt);
\draw[thick,blue] (10pt,3pt)--(55pt,3pt);

\draw[thick,blue] (-10pt,3pt)--(-55pt,3pt);

\draw[thick,blue] (10pt,-3pt)--(55pt,-3pt);

\draw[thick,blue] (-10pt,-3pt)--(-55pt,-3pt);

\draw[blue,thick] (10pt,3pt) arc (90:270:3pt);
\draw[blue,thick] (-10pt,3pt) arc (90:-90:3pt);
\draw[blue,thick] (55pt,3pt) arc (4:176:55pt);
\draw[blue,thick] (55pt,-3pt) arc (-4:-176:55pt);
\filldraw (7pt,0pt) circle (1pt) node[below]{\scriptsize $z_2$};
\filldraw (-7pt,0pt) circle (1pt) node[below]{\scriptsize $z_1$};
\filldraw (0pt,54pt) circle (1pt) node[below right]{\scriptsize $z_3$};
\filldraw (0pt,-54pt) circle (1pt) node[above right]{\scriptsize $z_4$};
\end{tikzpicture}
}
\hspace{30pt}
\subfloat[]{
\begin{tikzpicture}[scale=0.5,baseline={([yshift=0pt]current bounding box.center)}]
\draw [->,>=stealth] (-180pt,-50pt) -- (120pt,-50pt) node[right]{\scriptsize  $\Re(i\theta)$};
\draw [->,>=stealth] (0pt, -120pt) -- (0pt,140pt) node[right]{\scriptsize $\Im(i\theta)$};
\filldraw (0pt,-100pt) circle (2pt) node[right]{\scriptsize $\theta_4=it_4- \frac{\pi}{2}$};
\draw[thick,blue,->,>=stealth] (-2pt,-100pt)--(-2pt,-75pt);
\draw[thick,blue] (-2pt,-52pt)--(-2pt,-75pt);
\draw[thick,blue,->,>=stealth] (-2pt,-52pt)--(-100pt,-52pt);
\draw[thick,blue] (-150pt,-52pt)--(-75pt,-52pt);
\draw[thick,blue,-<,>=stealth] (-2pt,-48pt)--(-50pt,-48pt);
\draw[thick,blue] (-150pt,-48pt)--(-40pt,-48pt);
\filldraw (-150pt,-50pt) circle (2pt) node[above]{\scriptsize $\theta_2=i   t_2$};
\draw[thick,blue,->,>=stealth] (-2pt,-48pt)--(-2pt,-25pt);
\draw[thick,blue] (-2pt,0pt)--(-2pt,-25pt);
\filldraw (0pt,0pt) circle (2pt) node[right]{\scriptsize $ \theta_3=it_3+\frac{\pi}{2}$};
\draw[thick,blue,->,>=stealth] (-2pt,0pt)--(-2pt,25pt);
\draw[thick,blue] (-2pt,48pt)--(-2pt,25pt);
\draw[thick,blue,->,>=stealth] (-2pt,48pt)--(-100pt,48pt);
\draw[thick,blue] (-150pt,48pt)--(-75pt,48pt);
\draw[thick,blue,-<,>=stealth] (-2pt,52pt)--(-50pt,52pt);
\draw[thick,blue] (-150pt,52pt)--(-40pt,52pt);
\draw[thick,blue,->,>=stealth] (-2pt,75pt)--(-2pt,100pt);
\draw[thick,blue] (-2pt,52pt)--(-2pt,75pt);
\filldraw (-150pt,50pt) circle (2pt) node[above]{\scriptsize $\theta_1=i t_1+\pi$};
\end{tikzpicture}}
\caption{Operators $X_1(\theta_1),\ldots,X_4(\theta_4)$ of complex times $\theta_j=\tau_j+it_j$ on the double Keldysh contour, with $\tau_1,\ldots,\tau_4$ alternating and evenly spaced: (a)~The complex coordinate $z=e^{i\theta}$ is used so that points with positive real time $t$ are located inside the unit disk, $|z|=e^{-t}<1$; (b)~The contour is drawn such that the real time goes to the left, which is convenient when acting by operators on left. (In this example, $t_1=t_2$ and $t_3=t_4=0$.)}
\label{fig: OTO}
\end{figure}

For convenience, we set $\beta=2\pi$. Thus, the Lyapunov exponent is a dimensionless number, $0< \varkappa \leq 1$. We use the complex time $\theta =\tau+it$ and order operators according to $\tau$, i.e.\ the real part of $\theta$. In this paper, we consider the connected OTOC
\begin{equation}\label{OTOC0}
\langle X_1(\theta_1) X_2(\theta_2) \rangle \langle X_3(\theta_3) X_4(\theta_4) \rangle \mp
\langle X_1(\theta_1) X_3(\theta_3) X_2 (\theta_2) X_4 (\theta_4) \rangle
\end{equation}
with four complex times $\theta_{j} = \tau_j+ i  t_j$ satisfying the conditions
\begin{equation}
2\pi + \tau_4 \geq  \tau_1 \geq \tau_3 \geq \tau_2 \geq  \tau_4\,, \qquad
\frac{t_1+t_2}{2}-\frac{t_3+t_4}{2} \gg \kap^{-1}\,, \qquad
t_1-t_2 \sim t_3-t_4\sim 1\,.
\end{equation}
(The $-$ and $+$ signs in \eqref{OTOC0} are for bosons and fermions respectively.) See figure~\ref{fig: OTO} for an example of a configuration where the operators are evenly spaced on the imaginary time circle. 

\begin{defn}[Single-mode ansatz for early time OTOCs.] 
Following \cite{KS17-soft},
we assume that the connected OTOC has the following form:
\begin{equation}\label{OTOC1}
\langle X_1 X_2 \rangle \langle X_3 X_4 \rangle \mp \langle X_1 X_3 X_2 X_4 \rangle \approx \frac{e^{i\varkappa (\pi - \theta_1 -\theta_2 + \theta_3 + \theta_4)/2   }}{C} \VF^\R_{X_1,X_2} (\theta_1- \theta_2) \VF^\A_{X_3,X_4} (\theta_3-\theta_4)
\end{equation}
with $O(\lambda^2)$ accuracy, where $\lambda = C^{-1} e^{\varkappa t}$ and $C$ is large, for instance, in the SYK model $C \sim \frac{N}{\beta J}$ and in gravity $C\sim G_N^{-1}$. The retarded and advanced vertex functions $\VF^\R_{X_1,X_2}$, $\VF^\A_{X_3,X_4}$ have the same symmetry properties as time-ordered Euclidean Green functions. In particular,
\begin{equation}
\VF^\zeta_{Y,X}(\theta)=\VF^\zeta_{X,Y}(2\pi-\theta)
=\pm\VF^\zeta_{X,Y}(-\theta),\qquad \zeta=\R,\A.
\end{equation}
(The first equality follows from cyclic symmetry of the correlator with the other pair of operators equal to each other, whereas the second equality should be understood as a definition of $\VF^\zeta_{X,Y}(-\theta)$ because initially, the vertex functions are defined for $\theta=\tau+it$ with $0<\tau<2\pi$.) Other symmetry and positivity properties are obtained by Hermitian conjugation.
\end{defn}

\begin{defn}[Shorthands and conventions.]
For concreteness, we take $X_1=X_2=\chi_j$,\, $X_3=X_4=\chi_k$ to be Majorana operators in the SYK model. The subscripts of the vertex functions are dropped, and differences of time variables are abbreviated as $\theta_{jk} = \theta_j-\theta_k$. We focus on the OTOC dependence on real times and fix the imaginary times as in figure~\ref{fig: OTO}, namely,
\begin{equation}
\theta_1 = it_1 +\pi\,,\quad \theta_2=it_2\,,\quad \theta_3=it_3+\frac{\pi}{2}\,,\quad \theta_4=it_4-\frac{\pi}{2}\,.
\label{eq: real times}
\end{equation}
Using this notation, we define the function
\begin{equation}
\OTOC(t_1,t_2,t_3,t_4) :=
- \frac{1}{N}\calF(\theta_1,\theta_2,\theta_3,\theta_4) 
  \approx
  \frac{e^{\varkappa (t_1+t_2-t_3-t_4)/2   }}{{C}} \VF^\R(t_{12}) \VF^\A (t_{34}) \,,
\label{eqn: general form}
\end{equation}
where $\calF$ denotes the imaginary time ordered correlator normalized as in~\cite{MS16-remarks},
\begin{equation}
\frac{1}{N^2}\sum_{j,k} \bigl\langle\T
\chi_{j}(\theta_1) \chi_{j}(\theta_2) \chi_{k}(\theta_3) \chi_{k}(\theta_4)
\bigr\rangle
=G(\theta_1,\theta_2)G(\theta_3,\theta_4)
+\frac{1}{N}\calF(\theta_1,\theta_2,\theta_3,\theta_4)\,.
\end{equation}
We have also switched to real time arguments for the vertex functions.

Green function are defined with additional phase factors as is customary in condensed matter literature. In particular
\begin{equation}
G(\theta_1,\theta_2)=
-\bigl\langle\T\chi_j(\theta_1)\chi_j(\theta_2)\bigr\rangle,\quad
G^\R(t_1,t_2)=-i\theta(t_1-t_2)
\bigl\langle\{\chi_j(it_1),\chi_j(it_2)\}\bigr\rangle\,.
\end{equation}
\end{defn}

\begin{defn}[Kinetic equation and the retarded kernel.] 
 References~\cite{Kit.KITP.2,MS16-remarks,KS17-soft} derive a Bethe-Salpeter equation for $\calF$ on the Euclidean time circle,
\begin{equation}
 \calF= \calF_0 + K \calF \,,
 \label{eqn: kernel equation}
\end{equation}
and interpret it in terms of ladder diagrams. Specifically, $\calF$ is the sum of ladders with $0,1,2,\ldots$ rungs, antisymmetrized under $\theta_3\leftrightarrow\theta_4$, while $\calF_0$ is the antisymmetrized ``ladder'' with no rungs:
\begin{equation}
\calF_0(\theta_1,\theta_2,\theta_3,\theta_4) =
-\begin{tikzpicture}[baseline={([yshift=-3pt]current bounding box.center)}]
\draw[thick] (0pt,10pt)--(30pt,10pt);
\draw[thick] (0pt,-10pt)--(30pt,-10pt);
\node at (-5pt,10pt) {\scriptsize $1$};
\node at (-5pt,-10pt) {\scriptsize $2$};
\node at (35pt,10pt) {\scriptsize $3$};
\node at (35pt,-10pt) {\scriptsize $4$};
\end{tikzpicture}
+
\begin{tikzpicture}[baseline={([yshift=-3pt]current bounding box.center)}]
\draw[thick] (0pt,10pt)--(30pt,-10pt);
\draw[thick] (0pt,-10pt)--(30pt,10pt);
\node at (-5pt,10pt) {\scriptsize $1$};
\node at (-5pt,-10pt) {\scriptsize $2$};
\node at (35pt,10pt) {\scriptsize $3$};
\node at (35pt,-10pt) {\scriptsize $4$};
\end{tikzpicture}
= -G(\theta_1,\theta_3)G(\theta_2,\theta_4)
+G(\theta_1,\theta_4)G(\theta_2,\theta_3)\,.
\end{equation}
The integral kernel of the operator $K$ is defined as a product of two-point functions:
\begin{equation}
K(\theta_1,\theta_2,\theta_3,\theta_4) = 
-\begin{tikzpicture}[baseline={([yshift=-4pt]current bounding box.center)}]
\draw[thick] (0pt,15pt)--(40pt,15pt);
\draw[thick] (0pt,-15pt)--(40pt,-15pt);
\draw[thick] (40pt,-15pt)..controls (45pt,-7pt) and (45pt,7pt)..(40pt,15pt);
\draw[thick] (40pt,-15pt)..controls (35pt,-7pt) and (35pt,7pt)..(40pt,15pt);
\node at (-5pt,15pt) {\scriptsize $1$};
\node at (-5pt,-15pt) {\scriptsize $2$};
\node at (48pt,15pt) {\scriptsize $3$};
\node at (48pt,-15pt) {\scriptsize $4$};
\end{tikzpicture}
=  -J^2 (q-1) G (\theta_{13})  G ({\theta}_{24}) G(\theta_{34})^{q-2}\,.
\end{equation}
The operator product $K\calF$ in (\ref{eqn: kernel equation}) is the integral over two auxiliary points $\theta_5,\theta_6$ on the time circle, see figure~\ref{fig: kernel equation}~(a):
\begin{equation}
K\calF (\theta_1,\theta_2,\theta_3,\theta_4)
= \int_0^{2\pi} d\theta_5  \int_0^{2\pi}d\theta_6\,
K(\theta_1,\theta_2,\theta_5,\theta_6)
\calF (\theta_5,\theta_6,\theta_3,\theta_4)\,.
\end{equation}

\begin{figure}
[t]
\center
\subfloat[Time circle]{
\begin{tikzpicture}[scale=0.8, baseline={(current bounding box.center)}]
\draw[->,>=stealth] (-80pt,0pt) -- (80pt,0pt) node[right] {\scriptsize $\Re e^{i\theta}$};
\draw[->,>=stealth] (0pt,-80pt) -- (0pt,80pt) node[right]{\scriptsize $\Im e^{i\theta}$};
\draw (0pt,0pt) circle (60pt);
\draw[blue,thick] (0pt, 0pt) circle (54pt);
\filldraw (54pt,0pt) circle (1pt) node[below left]{\scriptsize $2$};
\filldraw (-54pt,0pt) circle (1pt) node[above right]{\scriptsize $1$};
\filldraw (0pt,54pt) circle (1pt) node[below right]{\scriptsize $3$};
\filldraw (0pt,-54pt) circle (1pt) node[above right]{\scriptsize $4$};
\filldraw[thick, red, fill=white] (50pt,20pt) circle (1pt) node[left]{\scriptsize $5$};
\filldraw[thick, red, fill=white] (-45pt,-30pt) circle (1pt) node[right]{\scriptsize $6$};
\end{tikzpicture}
}
\hspace{30pt}
\subfloat[Deformed (double Keldysh) contour]{\hspace{15pt}
\begin{tikzpicture}[scale=0.8, baseline={(current bounding box.center)}]
\filldraw[gray, opacity=0.5] (-47pt,-5pt) rectangle  (-12pt,5pt) ;
\filldraw[gray, opacity=0.5] (47pt,5pt) rectangle  (12pt,-5pt) ;
\draw[->,>=stealth] (-80pt,0pt) -- (80pt,0pt) node[right] {\scriptsize $\Re e^{i\theta}$};
\draw[->,>=stealth] (0pt,-80pt) -- (0pt,80pt) node[right]{\scriptsize $\Im e^{i\theta}$};
\draw (0pt,0pt) circle (60pt);
\draw[ dashed, red] (0pt,0pt) circle (30pt);
\draw[thick,blue] (10pt,3pt)--(55pt,3pt);
\draw[thick,blue] (-10pt,3pt)--(-55pt,3pt);
\draw[thick,blue] (10pt,-3pt)--(55pt,-3pt);
\draw[thick,blue] (-10pt,-3pt)--(-55pt,-3pt);
\draw[blue,thick] (10pt,3pt) arc (90:270:3pt);
\draw[blue,thick] (-10pt,3pt) arc (90:-90:3pt);
\draw[blue,thick] (55pt,3pt) arc (4:176:55pt);
\draw[blue,thick] (55pt,-3pt) arc (-4:-176:55pt);
\filldraw (7pt,0pt) circle (1pt) node[below]{\scriptsize $2$};
\filldraw (-7pt,0pt) circle (1pt) node[below]{\scriptsize $1$};
\filldraw (0pt,54pt) circle (1pt) node[below right]{\scriptsize $3$};
\filldraw (0pt,-54pt) circle (1pt) node[above right]{\scriptsize $4$};
\filldraw[thick, red,fill=white] (30pt,3pt) circle (1pt);
\filldraw[thick, red,fill=white] (30pt,-3pt) circle (1pt);
\filldraw[thick, red,fill=white] (-30pt,3pt) circle (1pt);
\filldraw[thick, red,fill=white] (-30pt,-3pt) circle (1pt);
\end{tikzpicture}
\hspace{15pt}}
\caption{Integration contours for the Bethe-Salpeter equation \eqref{eqn: kernel equation}. The deformed contour (b) is suited for the OTOC calculation, where the main contribution to the integral comes from the folds shown in gray. There are 4 points (red) on the folds for each value of real time, $t=\Im\theta$.}
\label{fig: kernel equation}
\end{figure}

The Bethe-Salpeter equation \eqref{eqn: kernel equation} also holds on a deformed contour. The double Keldysh contour in figure~\ref{fig: kernel equation}~(b) is suited for the calculation of $\OTOC(t_1,t_2,t_3,t_4)$ related to $\calF$ by \eqref{eqn: general form}. We consider this function with $t_3,t_4$ fixed and denote it by $F(t_1,t_2)$. Since OTOCs grow exponentially, the $K\calF$ term is much greater than $\calF_0$, and hence, the latter may be neglected. Furthermore, the integration contour may be reduced to the union of two folds, marked as gray area in figure~\ref{fig: kernel equation}~(b), which are then extended to $t=-\infty$. Written in terms of real times, the equation is reduced to the following form:
\begin{equation}
\begin{gathered}
\int_{\text{folds}} d\theta_5\,d\theta_6\,
K(it_1+\pi,it_2,\theta_5,\theta_6) F (t_5,t_6) \approx F(t_1,t_2)\,,\\
\theta_5=it_5,\,it_5+\pi,\quad\, \theta_6=it_6,\,it_6+\pi.
\end{gathered}
\end{equation}
For each value of $t_5$, there are four points on the contour, two on the right (with $\theta_5=it_5$) and two on the left (with $\theta_5=it_5+i\pi$). The contributions from the first two points cancel as the corresponding sections of the contour are traversed in opposite directions and the integrand is the same. Note, however, that the kernel $K$ contains the contour-ordered Green function $G(\theta_1,\theta_5)$ which is different at the other two points; the difference is equal to $-iG^{\R}(t_1,t_5)$. The same consideration is applicable to $\theta_6$. Thus, we obtain a kinetic equation for the OTOC,
\begin{equation}
 \int_{\RR} dt_5\,dt_6\, K^\R (t_1,t_2,t_5,t_6) F (t_5,t_6)
\approx F (t_1,t_2)\,,
 \label{eq: kinetic equation}
\end{equation}
with the retarded kernel
\begin{equation}
K^\R(t_1,t_2,t_5,t_6) = 
- \begin{tikzpicture}[baseline={([yshift=-4pt]current bounding box.center)}]
\draw[thick] (0pt,15pt)--(40pt,15pt);
\draw[thick] (0pt,-15pt)--(40pt,-15pt);
\node at (20pt,20pt) {\scriptsize $\R$};
\node at (20pt,-20pt) {\scriptsize $\R$};
\node at (51pt,0pt) {\scriptsize $\W$};
\draw[thick] (40pt,-15pt)..controls (45pt,-7pt) and (45pt,7pt)..(40pt,15pt);
\draw[thick] (40pt,-15pt)..controls (35pt,-7pt) and (35pt,7pt)..(40pt,15pt);
\node at (-5pt,15pt) {\scriptsize $1$};
\node at (-5pt,-15pt) {\scriptsize $2$};
\node at (48pt,15pt) {\scriptsize $5$};
\node at (48pt,-15pt) {\scriptsize $6$};
\end{tikzpicture}
=  - J^2 (q-1) G^\R (t_1,t_5)  G^\R (t_2,t_6) G^\W(t_5,t_6)^{q-2}\,,
\label{retarded_kernel}
\end{equation}
where $G^\R$ is the retarded Green function and $G^\W(t,t')=-\langle\chi_j(it+\pi)\chi_j(it')\rangle$ is the Wightman function with $\pi$ separation in the imaginary time. The retarded kernel $K^\R$ is invariant under the translation of all four times. Let us also define a variant of the kernel with a parameter $\alpha<0$:
\begin{equation}
K^\R_{\alpha} (t,t')= \int {K}^\R
\biggl(s+ \frac{t}{2},\, s- \frac{t}{2},\, \frac{t'}{2},\, -\frac{t'}{2}\biggr)
\,e^{\alpha s}\, ds \,.
\label{eqn: K kappa}
\end{equation}
The kernel $K^\R_{\alpha}$ represents an operator acting on functions of $t'$, and its largest eigenvalue is denoted by $k_\R(\alpha)$. Finding the Lyapunov exponent amounts to solving the equation ${k_\R(-\varkappa)=1}$.\,\footnote{In the conformally invariant case, our definition of the function $k_\R$ is the same as in~\cite{MS16-remarks}.}
\end{defn}

\begin{defn}[Nonlinearity at later times.] As was previously mentioned, we focus on the OTOC term that is linear in $\lambda=C^{-1}e^{\kap t}$. Although late time OTOCs are beyond the scope of this paper, some applications require a qualitative understanding of nonlinear effects. Corrections of order $\lambda^m$ to the four-point correlator are given by $m$ parallel ladders that are joined together in a tree-like fashion. This is a schematic drawing of the sum of all such diagrams for $m=3$, with the ladders depicted as wavy lines: 
\begin{equation}
\begin{tikzpicture}
 [baseline={([yshift=-4pt]current bounding box.center)},
  vertex/.style={circle,draw=black,thick,fill=lightgray,minimum size=16pt},
  wavy/.style={thick,decorate,decoration={snake,amplitude=2pt,segment length=5pt}}]
\node[vertex] (R) at (-30pt,0pt) {};
\node[vertex] (A) at (30pt,0pt) {};
\draw[thick] (R) -- ++(135:20pt) node[left]{\scriptsize$1$};
\draw[thick] (R) -- ++(-135:20pt) node[left]{\scriptsize$2$};
\draw[thick] (A) -- ++(45:20pt) node[right]{\scriptsize$3$};
\draw[thick] (A) -- ++(-45:20pt) node[right]{\scriptsize$4$};
\draw[wavy] (A) to[out=135,in=45] (R);
\draw[wavy] (A) to (R);
\draw[wavy] (A) to[out=-135,in=-45] (R);
\end{tikzpicture}\,.
\end{equation}
Based on this picture, we generalize the OTOC ansatz as follows:
\begin{equation}
-\frac{1}{N^2}\sum_{j,k}\bigl\langle \chi_{j}(\theta_1) 
\chi_{k}(\theta_3)
\chi_{j}(\theta_2)
\chi_{k}(\theta_4)  \bigr\rangle
\approx \sum_{m=0}^{\infty}\frac{(-\lambda)^m}{m!}
\VF^{\R,m}(t_{12})\VF^{\A,m}(t_{34})\,,
\label{eq: nlOTOC}
\end{equation}
where $\theta_1,\theta_2,\theta_3,\theta_4$ are defined in \eqref{eq: real times} and $\lambda=C^{-1}e^{\kap(t_1+t_2-t_3-t_4)/2}$. In particular, $\VF^{\R,0}$ and $\VF^{\A,0}$ are equal to the Wightman function $G^\W$, and $\VF^{\R,1}=\VF^\R$,\, $\VF^{\A,1}=\VF^A$ are the vertex functions considered previously. For many purposes, the dependence of the OTOC \eqref{eq: nlOTOC} on $t_1,t_2$ is well represented by a non-equilibrium analogue of the Wightman function,
\begin{equation}
F_{c}(t_1,t_2)=\sum_{m=0}^{\infty}
\frac{\bigl(c\,e^{\kap(t_1+t_2)/2}\bigr)^m}{m!}\VF^{\R,m}(t_{12})\,,
\end{equation}
which satisfies a certain nonlinear equation for any value of the initial perturbation strength $c$. Nonlinear kinetic equations on the double Keldysh contour were studied in~\cite{AFI2016}; they generally have $G^\W$ as an unstable fixed point and $0$ as a stable fixed point.
\end{defn}

\section{The ladder identity and its derivation}
\begin{defn}[The ladder identity.] 
The following identity holds for the SYK model:
\begin{equation}
\frac{2 \cos \frac{\varkappa \pi}{2}}{\tilde{C}} \cdot k'_\R(-\varkappa) =1\,, \quad \text{where} \quad \tilde{C} = \frac{C}{N\left( \VF^\A, \VF^\R \right) } \,.
\label{eq: identity}
\end{equation}
The notation $\left( \VF^\A,\VF^\R\right)$ stands for the inner product of vertex functions:
\begin{equation}\label{inner_product}
\left( \VF^\A, \VF^\R \right) := (q-1) J^2 \int dt\, \VF^\A(t) \bigl(G^\W(t) \bigr)^{q-2} \VF^\R (t)\,.
\end{equation} 
The second factor in (\ref{eq: identity}) is the derivative of the eigenvalue $k_\R(\alpha)$ at $\alpha=-\varkappa$.  
It has the dimension of time and will be called \emph{branching time}:
\begin{equation}
t_B:= k_\R'(-\varkappa)\,.
\end{equation}
We can express the branching time $t_B$ using the following formula:
\begin{gather}
\label{eqn: branching}
t_B = \frac{1}{\left(\VF^\A,\VF^\R\right)}\int \BOX
\biggl(s+\frac{t}{2},\, s-\frac{t}{2},\, \frac{t'}{2},\, -\frac{t'}{2}\biggr)\,
\VF^\A(t) \VF^\R(t')
e^{-\varkappa s} s\, ds\, dt\, dt' \,,
\\[8pt]
\label{eqn: box}
\begin{aligned}
\BOX(t_1,t_2,t_3,t_4) &=
- \begin{tikzpicture}[baseline={([yshift=-4pt]current bounding box.center)}]
\draw[thick] (0pt,-15pt)..controls (5pt,-7pt) and (5pt,7pt)..(0pt,15pt);
\draw[thick] (0pt,-15pt)..controls (-5pt,-7pt) and (-5pt,7pt)..(0pt,15pt);
\draw[thick] (0pt,15pt)--(40pt,15pt);
\draw[thick] (0pt,-15pt)--(40pt,-15pt);
\draw[thick] (40pt,-15pt)..controls (45pt,-7pt) and (45pt,7pt)..(40pt,15pt);
\draw[thick] (40pt,-15pt)..controls (35pt,-7pt) and (35pt,7pt)..(40pt,15pt);
\node at (-11pt,0pt) {\scriptsize $\W$};
\node at (20pt,20pt) {\scriptsize $\R$};
\node at (20pt,-20pt) {\scriptsize $\R$};
\node at (51pt,0pt) {\scriptsize $\W$};
\node at (-8pt,15pt) {\scriptsize $1$};
\node at (-8pt,-15pt) {\scriptsize $2$};
\node at (48pt,15pt) {\scriptsize $3$};
\node at (48pt,-15pt) {\scriptsize $4$};
\end{tikzpicture}\\
&= - \bigl(J^2 (q-1)\bigr)^2 G^\W(t_1,t_2)^{q-2} G^\R (t_1,t_3) G^\R(t_2,t_4) G^\W(t_3,t_4)^{q-2}\,.
\end{aligned}
\end{gather}
More explicitly, this is done by the following steps:
\begin{enumerate}
\item  Denote the eigenfunction of $K_\alpha^\R$ by $\VF^\R_{\alpha}$:
\begin{equation}
K_\alpha^\R\VF^\R_{\alpha} = k_\R(\alpha)\, \VF^\R_{\alpha}\,.
\label{eqn: eigen equation}
\end{equation} 
Similarly, $\VF^\A_{\alpha}$ is the eigenfunction (with the same eigenvalue) of the operator $K_\alpha^\A$ that is adjoint to $K_\alpha^\R$ with respect to the inner product \eqref{inner_product}. When the eigenvalue $k_\R(\alpha)$ is $1$, i.e. when $\alpha=-\varkappa$, the eigenfunctions $\VF^\R_{\alpha},\VF^\A_{\alpha}$ are, respectively, the retarded and advanced vertex functions $\VF^\R,\VF^\A$ defined in (\ref{eqn: general form}).
\item Multiply (\ref{eqn: eigen equation}) by $\VF^\A_\alpha$ on the left: \begin{equation}
\bigl(\VF^\A_{\alpha}, K^\R_{\alpha}\VF^\R_{\alpha}\bigr)=
k_\R(\alpha) \bigl( \VF^\A_{\alpha}, \VF^\R_{\alpha} \bigr)\,.
\end{equation}
\item Take the derivative with respect to $\alpha$. The left-hand side of the above equation becomes
\begin{equation}
\biggl( \frac{d\VF^\A_\alpha}{d\alpha},\, K^\R_{\alpha}\VF^\R_\alpha \biggr)
+\biggl( \VF^\A_{\alpha},\, \frac{dK^\R_{\alpha}}{d\alpha} \VF^\R_{\alpha} \biggr)
+\biggl( \VF^\A_{\alpha},\, K^\R_{\alpha}\frac{d\VF^\R_{\alpha}}{d\alpha} \biggr) \,.
\end{equation}
Adding the first and third terms, we get
$
k_\R(\alpha)\,\frac{d}{d\alpha}\bigl(\VF^\A_{\alpha},\VF^\R_{\alpha}\bigr)
$, 
which cancels the corresponding term on the right-hand side. In the resulting equation, we set $\alpha=-\varkappa$ and recall that $\frac{dk_\R(\alpha)}{d\alpha}\big|_{\alpha=-\kap}=t_B$. Thus,
\begin{equation}
\biggl( \VF^\A_{\alpha},\, \frac{dK^\R_{\alpha}}{d\alpha}\bigg|_{\alpha=-\kap} \VF^\R_{\alpha} \biggr)
= t_B \bigl(\VF^\A, \VF^\R\bigr)\,.
\end{equation}
Inserting the explicit definition (\ref{eqn: K kappa}) of $K^\R_{\alpha}$, we get a formula equivalent to (\ref{eqn: branching}).
\end{enumerate}
\end{defn}

\begin{defn}[Derivation of the ladder identity.] OTOCs in the SYK model are sums of ladder diagrams. When the times are well separated, $\frac{t_1+t_2-t_3-t_4}{2}\gg\kap^{-1}$, the sum is dominated by sufficiently long ladders. The idea is to cut them into smaller pieces and find a consistency condition. We consider the connected OTOC and express it using the ansatz:
\begin{equation}
\OTOC(t_1,t_2,t_3,t_4) \approx \frac{e^{ \varkappa ( t_1 + t_2 - t_3 - t_4)/{2}}}{C} \VF^\R( t_{12}) \VF^\A(t_{34}) \,.
\label{eqn: whole diagram}
\end{equation}
Alternatively, $\OTOC(t_1,t_2,t_3,t_4)$ can be computed by dividing each contributing ladder into three parts as shown in figure~\ref{fig: identity derivation}(a). The computation is set up as follows:

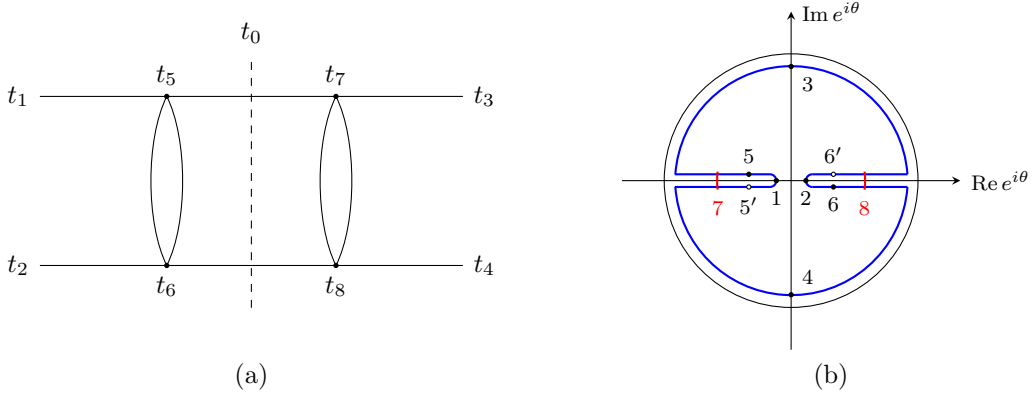
\begin{figure}[t]
\center
\subfloat[]{\begin{tikzpicture}[scale=0.8]
\draw (-100pt,-40pt) -- (100pt,-40pt);
\draw (-100pt,40pt) -- (100pt,40pt);
\node[left] at (-100pt,40pt) {\footnotesize $t_1$};
\node[left] at (-100pt,-40pt) {\footnotesize $t_2$};
\node[right] at (100pt,40pt) {\footnotesize $t_3$};
\node[right] at (100pt,-40pt) {\footnotesize $t_4$};
\draw[dashed] (0pt,-60pt)--(0pt,60pt) node[above] {\footnotesize $t_0$};
\filldraw (-40pt,40pt) circle (1pt) node[above] {\footnotesize $t_5$};
\filldraw (-40pt,-40pt) circle (1pt) node[below] {\footnotesize $t_6$};
\filldraw (40pt,40pt) circle (1pt) node[above] {\footnotesize $t_7$};
\filldraw (40pt,-40pt) circle (1pt) node[below] {\footnotesize $t_8$};
\draw (-40pt,-40pt) ..controls (-50pt,-20pt) and (-50pt,20pt) .. (-40pt,40pt) ;
\draw (-40pt,-40pt) ..controls (-30pt,-20pt) and (-30pt,20pt) .. (-40pt,40pt) ;
\draw (40pt,-40pt) ..controls (50pt,-20pt) and (50pt,20pt) .. (40pt,40pt) ;
\draw (40pt,-40pt) ..controls (30pt,-20pt) and (30pt,20pt) .. (40pt,40pt) ;
\path (-100pt,-80pt) -- (100pt,-80pt);
\end{tikzpicture}}
\hspace{40pt}
\subfloat[]{\begin{tikzpicture}[scale=0.8]
\draw[->,>=stealth] (-80pt,0pt) -- (80pt,0pt) node[right] {\scriptsize $\Re e^{i\theta}$};
\draw[->,>=stealth] (0pt,-80pt) -- (0pt,80pt) node[right]{\scriptsize $\Im e^{i\theta}$};
\draw (0pt,0pt) circle (60pt);

\draw[thick,blue] (10pt,3pt)--(55pt,3pt);
\draw[thick,blue] (-10pt,3pt)--(-55pt,3pt);
\draw[thick,blue] (10pt,-3pt)--(55pt,-3pt);
\draw[thick,blue] (-10pt,-3pt)--(-55pt,-3pt);

\draw[blue,thick] (10pt,3pt) arc (90:270:3pt);
\draw[blue,thick] (-10pt,3pt) arc (90:-90:3pt);
\draw[blue,thick] (55pt,3pt) arc (4:176:55pt);
\draw[blue,thick] (55pt,-3pt) arc (-4:-176:55pt);

\filldraw (7pt,0pt) circle (1pt) node[below]{\scriptsize $2$};
\filldraw (-7pt,0pt) circle (1pt) node[below]{\scriptsize $1$};
\filldraw (0pt,54pt) circle (1pt) node[below right]{\scriptsize $3$};
\filldraw (0pt,-54pt) circle (1pt) node[above right]{\scriptsize $4$};

\filldraw[fill=white] (20pt,3pt) circle (1pt) node[above]{\scriptsize $6'$};
\filldraw[fill=white] (-20pt,-3pt) circle (1pt) node[below]{\scriptsize $5'$};

\filldraw[fill=black] (20pt,-3pt) circle (1pt) node[below]{\scriptsize $6$};
\filldraw[fill=black] (-20pt,3pt) circle (1pt) node[above]{\scriptsize $5$};

\draw[red,thick] (0pt,0pt) ++(173:35pt) arc (173:187:35pt) node[below]{\scriptsize $7$};
\draw[red,thick] (0pt,0pt) ++(7:35pt) arc (7:-7:35pt) node[below]{\scriptsize $8$};
\end{tikzpicture}}
\caption{(a)~Dividing each ladder diagram into three parts: the central box of width ${s= \frac{t_5+t_6-t_7-t_8}{2}}$, an OTOC of times $t_1,t_2,t_5,t_6$, and an OTOC of times $t_7,t_8,t_3,t_4$;\, (b)~The location of points $7$ and $8$ on the folds and two choices for $5$ and $6$.}
\label{fig: identity derivation}
\end{figure}

\begin{enumerate}
\item Pick a time $t_0$ away from the boundary, such that $\frac{t_1+t_2}{2}-t_0,\, t_0-\frac{t_3+t_4}{2}\gg \kap^{-1}$. Find the adjacent rungs $(t_5,t_6)$ and $(t_7,t_8)$ satisfying the condition
\begin{equation}
 \frac{t_5+t_6}{2} > t_0> \frac{t_7+t_8}{2} \,.
\label{eqn: constraint}
\end{equation}
The ladder diagram contains the factor $\BOX(t_5,t_6,t_7,t_8)$ as in \eqref{eqn: box}. (However, until we sum over the positions of points on the folds, all Green functions are contour-ordered.) The factors corresponding to the left and right parts of the ladder are denoted by $\OTOC_A$ and $\OTOC_B$, respectively. The contour integral over points $5,6,7,8$ is multiplied by $1/2$ because both $\OTOC_A$ and $\OTOC_B$ are antisymmetrized, whereas the result should be antisymmetrized only once.

\item For the arguments of $\OTOC_B$ to be out of time order (so that the correlator is relatively large), the points $7$ and $8$ must be separated by $3$ and $4$ on the double Keldysh contour. Without loss of generality, we may assume that $7$ is on the left fold and $8$ is on the right fold; this also removes the $1/2$ factor. As in the derivation of the kinetic equation, $5$ should be on the same fold as $7$, and $6$ on the same fold as $8$. Summing over the positions of points $7$ and $8$ with $5$ and $6$ fixed, we recover equation \eqref{eqn: box} for the box with retarded Green functions.

\item Imposing the requirement that the arguments of $\OTOC_A$ are out of time order, we are left with two choices for points $5$ and $6$ shown in figure~\ref{fig: identity derivation}(b):\, $(5,6)$ or $(5',6')$. In the first case,
\begin{equation}
\OTOC_A =
\OTOC\biggl(t_1,t_2,\, t_5-i\frac{\pi}{2},\,t_6-i\frac{\pi}{2}\biggr)
\approx \frac{ e^{\varkappa (i\pi+t_1+t_2-t_5-t_6)/2} }{C}
\VF^\R(t_{12}) \VF^\A(t_{56})\,.
\label{eqn: OTOCA}
\end{equation}
The shift in the time arguments is because $\theta_1,\theta_2,\theta_5,\theta_6$ are not evenly spaced on the imaginary time circle, namely, $\theta_5=it_5+\pi$ and $\theta_6=it_6$ (instead of $it_5+\frac{\pi}{2}$ and $it_6-\frac{\pi}{2}$). If $5'$ and $6'$ are chosen, then
\begin{equation}
\OTOC_{A'} =
-\OTOC\biggl(t_1,t_2,\, t_6+i\frac{\pi}{2},\,t_5+i\frac{\pi}{2}\biggr)
\approx \frac{e^{\varkappa (-i \pi +t_1+t_2-t_5-t_6)/{2}}}{C}
\VF^\R(t_{12}) \VF^\A(t_{56})\,,
\label{eqn: OTOCA'}
\end{equation}
where we have used the antisymmetry of the four-point correlator and the advanced vertex function $\VF^\A(t_{56})$. In both cases,
\begin{equation}
\OTOC_B \approx \frac{ e^{\varkappa (t_7+t_8-t_3-t_4)/2}}{C} \VF^\R(t_{78}) \VF^\A(t_{34}) \,.
\end{equation} 
\end{enumerate}

Now, $\OTOC(t_1,t_2,t_3,t_4)$ can be expressed as a product, $(\OTOC_A+\OTOC_{A'}) \cdot \BOX \cdot \OTOC_B$; the exact equality holds if all the OTOCs are multiplied by $N$. Thus,
\begin{align}
\OTOC(t_1,t_2,t_3,t_4) \approx{}
& N \, \frac{2\cos \frac{\varkappa \pi}{2}}{  C} \,
\frac{ e^{ \varkappa (t_1 + t_2 - t_3 - t_4)/{2}} }{C}\,
\VF^\R(t_{12}) \VF^\A(t_{34}) 
\label{eqn: final integral}\\
& \cdot\int dt_5\,dt_6\,dt_7\,dt_8\,
\VF^\A(t_{56})\BOX(t_5,t_6,t_7,t_8)\,\VF^\R(t_{78})\,
e^{-\varkappa (t_5+t_6-t_7-t_8)/{2}}\,.\nn
\end{align}
For convenience, we switch to new integration variables $s,t_*, t_{56},t_{78}$, where
\begin{equation}
s = \frac{t_5+t_6-t_7-t_8}{2} ,\quad t_* = \frac{t_5+t_6+t_7+t_8}{4} \,.
\end{equation}
The integrand of (\ref{eqn: final integral}) does not depend on $t_*$, which is constrained to an interval of length $s$ due to \eqref{eqn: constraint}, namely, $t_0-\frac{s}{2}<t_{*}<t_0+\frac{s}{2}$. Therefore, we have the following formula for the OTOC:
\begin{align}
\OTOC(t_1,t_2,t_3,t_4) \approx{}
& N \, \frac{2\cos \frac{\varkappa \pi}{2}}{  C} \,
\frac{ e^{ \varkappa (t_1 + t_2 - t_3 - t_4)/{2}} }{C}\,
\VF^\R(t_{12}) \VF^\A(t_{34}) 
\\
& \cdot\int s\,ds\,dt_{56}\,dt_{78}\,
\BOX\biggl(s+\frac{t_{56}}{2},\, s-\frac{t_{56}}{2},\,
\frac{t_{78}}{2},\, -\frac{t_{78}}{2}\biggr)\,
\VF^\A(t_{56})\VF^\R(t_{78})\,e^{-\varkappa s}\,.\nn
\end{align}
The last line is equal to $\bigl(\VF^\A,\VF^\R\bigr)t_B$ due to \eqref{eqn: branching}, and the left-hand side may be written in the form (\ref{eqn: whole diagram}). Thus we have obtained the identity (\ref{eq: identity}).
\end{defn}

Here we also remark on the meaning of the branching time $t_B$. The integrand of (\ref{eqn: branching}) includes the factor $s= \frac{t_5+t_6}{2} - \frac{t_7+t_8}{2}$, the distance between two rungs of the ladder, whereas the remaining part may be interpreted as a statistical weight; therefore, the formula has the meaning of the average rung separation.

\section{Applications}

\subsection{Computational shortcuts}

Thanks to the ladder identity, it is sufficient to compute one of the numbers $C$ and $\kap$; the other one is obtained almost automatically. We will illustrate this point by two calculations for the SYK model,
\begin{equation}
H= i^{\frac{q}{2}} \sum_{1\leq j_1 \ldots < j_q \leq N} J_{j_1,\ldots,j_q} \chi_{j_1}  \ldots \chi_{j_q}, \qquad \{\chi_j,\chi_k\}=\delta_{jk},\qquad
\bar{J^2_{j_1,\ldots , j_q}} = \binom{N-1}{q-1}^{-1} J^2\,,
\end{equation}
in different regimes. We assume that $N$ is very large, i.e.\ take the $N\to\infty$ limit before any other limit. The notation $\Delta=1/q$ is used; also recall that $\beta=2\pi$.

\begin{defn}[Prefactor from the retarded kernel.]
Solving the kinetic equation (\ref{eq: kinetic equation}), one obtains the Lyapunov exponent $\varkappa$ and vertex functions $\VF^\R,\VF^A$. Although the prefactor $C^{-1}$ of the OTOC is not determined by the kinetic equation itself, it can be expressed using the ladder identity,
\begin{equation}
N\,\frac{2\cos\frac{\varkappa \pi}{2}}{C}
k'_\R(-\varkappa)\bigl(\VF^\A,\VF^\R\bigr)=1\,.
\end{equation}
For example, let us calculate the OTOC for the SYK model at $q\to\infty$ while keeping $\calJ=\sqrt{2^{1-q}q}\,J$ fixed. The large $q$ limit was studied in \cite{MS16-remarks}, where the Lyapunov exponent and some other quantities were computed for an arbitrary $\calJ$. In particular, the imaginary-time Green function is
\begin{equation}
G(\tau) = -\frac{1}{2} \sgn(\tau)
\left[ \frac{\cos \frac{\pi v}{2} }{ \cos \bigl(\pi v
\bigl(\frac{1}{2}-\frac{|\tau|}{2 \pi}\bigr)\bigr)}  \right]^{2\Delta},\qquad
\frac{v}{2\cos\frac{\pi v}{2}} =\calJ
\end{equation}
(with $O(q^{-2})$ accuracy), where the parameter $v\in (0,1)$ characterizes the coupling strength. The retarded kernel, its eigenfunctions, and eigenvalue have the following explicit form:
\begin{equation}
K^\R =\theta(t_{13})\theta(t_{24})\frac{v^2}{2\cosh^2\frac{v t_{34}}{2}}\,,\quad
\VF^\R_{\alpha}(t) = \VF^\A_{\alpha}(t) =
\biggl(2\cosh\frac{vt}{2}\biggr)^{\frac{\alpha}{v}},\quad
k_\R(\alpha) = \frac{2v^2}{\alpha(\alpha-v)} \,.
\end{equation}
The Lyapunov exponent is determined by the equation $k_\R(-\varkappa)=1$, which has a solution ${\varkappa=v}$. The branching time and the inner product between the eigenfunctions are computed easily:
\begin{equation}
t_B = k'_\R (-v)= \frac{3}{2v}\,,\quad
\bigl(\VF^\A, \VF^\R\bigr)=\frac{v}{3}\,.
\end{equation}
Applying the identity, we get the coefficient $C$:
\begin{equation}
C =N \cdot 2\cos \frac{\varkappa \pi}{2} \cdot t_B \cdot\left( \VF^\A, \VF^\R \right)
=N \cos \frac{v\pi}{2} \,.
\end{equation}
Thus, the OTOC for the large $q$ SYK is given by
\begin{equation}
\text{OTOC}(t_1,t_2,t_3,t_4) \approx \frac{1  }{N \cos \frac{v\pi}{2}} \frac{e^{v(t_1+t_2-t_3-t_4)/{2}} }{\left(2\cosh \frac{vt_{12}}{2} \right) \left(2 \cosh \frac{vt_{34}}{2} \right)} \,.
\end{equation}
This result is consistent with that in~\cite{Qi_Streicher} at $t\gg 1$, where an operator growth picture is used in the computation.
\end{defn}

\begin{defn}[Near-maximal chaos.] Now, we consider the SYK model at a fixed $q>2$ and $J\to\infty$. In this limit, $\kap$ becomes arbitrarily close to $1$, which may be explained by an emergent $\SL(2,\RR)$ symmetry. The symmetry is manifest in the retarded and Wightman Green functions,
\begin{equation}
G^\R(t_1,t_2)= -i\theta(t_{12})\frac{2 b^{\Delta} \cos \pi \Delta }{ \left( 2 J \sinh \frac{t_{12}}{2} \right)^{2\Delta}}\,, \qquad
G^\W(t_1,t_2)= -\frac{b^\Delta}{\left( 2 J\cosh \frac{t_{12}}{2} \right)^{2\Delta}}\,, 
\end{equation}
where $b=\frac{1}{\pi}\bigl(\frac{1}{2}-\Delta\bigr) \tan(\pi\Delta)$. Indeed, these functions are invariant under the simultaneous action of the symmetry generators
\begin{equation}
L_m=e^{-mt}(\partial_t-m\Delta),\quad m=-1,0,1
\end{equation}
on both time variables. Acting by $L_m$ on the first variable of the Wightman function, we get some functions that are invariant under both the retarded and the advanced kernels. In particular, $L_{-1}$ and $L_{1}$ generate $\tilde{\VF}^\R(t_1,t_2) =e^{(t_1+t_2)/2}\VF^\R(t_{12})$ and $\tilde{\VF}^\A(t_1,t_2) =e^{-(t_1+t_2)/2}\VF^\A(t_{12})$, respectively, where
\begin{equation}
\VF^\R(t)=\VF^\A(t)=-\frac{2\Delta b^{\Delta}J^{-2\Delta}}
{\bigl(2\cosh\frac{t}{2}\bigr)^{2\Delta+1}}\,.
\end{equation}
The low energy dynamics of the model is described by the Schwarzian theory\cite{Kit.KITP.2,MS16-remarks,KS17-soft},
\begin{equation}
I_{\Sch}[\varphi] = - \frac{N \alpha_S}{J} \int_0^{2\pi}  \Sch \left( e^{i\varphi(\tau)} ,\tau \right) d\tau \,,\quad
\varphi\in\operatorname{Diff}^{+}(S^1)\,,
\end{equation}
which allows for the calculation of the OTOC prefactor~\cite{MS16-remarks,MSY16,KS17-soft}:
\begin{equation}
C=\frac{2N\alpha_S}{J}\,.
\end{equation}

However, the above analysis is not exact. In fact, the Lyapunov exponent has a small correction, $\dkap:=1-\kap\sim J^{-1}$, which is essential for commutator OTOCs (see section~\ref{sec: stringy}). We now streamline the difficult calculation of $\dkap$~\cite{MS16-remarks} by using the ladder identity. The eigenfunction and eigenvalue of the retarded kernel are as follows~\cite{MS16-remarks}:
\begin{equation}
\VF^\R_\alpha(t)
\propto\biggl(2\cosh\frac{t}{2}\biggr)^{-2\Delta+\alpha},\qquad
k_\R(\alpha) = \frac{\Gamma(3-2\Delta)\Gamma(2\Delta-\alpha)}
{\Gamma(1+2\Delta) \Gamma(2-2\Delta-\alpha) }\,.
\end{equation}
(To check the eigenvalue equation, it is convenient to pass to the variables $z_1=-e^{-t_1}$ and $z_2=e^{-t_2}$ so that the function $\tilde{\VF}^\R_\alpha(t_1,t_2) =e^{-\alpha(t_1+t_2)/2} \VF^\R_\alpha(t_{12})$ transformed as a $\Delta$-form in each variable becomes $(z_2-z_1)^{-2\Delta+\alpha}$.) Hence,
\begin{equation}
t_B = k'_\R(-1) = 
\pi\cot(2\pi\Delta)
-\frac{1}{2\Delta}-\frac{1}{2\Delta-1}-\frac{1}{2\Delta-2}\,.
\end{equation}
The branching time $t_B$ reaches its maximum $t_B=\frac{3}{2}$ at $\Delta=0$ and minimum $t_B=0$ at $\Delta=\frac{1}{2}$. The inner product $(\VF^\A,\VF^\R)$ can also be computed explicitly:
\begin{equation}
\bigl(\VF^\A,\VF^\R\bigr)
=\frac{\Delta(1-\Delta)(1-2\Delta)}{3\pi}\tan(\pi\Delta)\,.
\end{equation}
Applying the ladder identity, we approximate the factor $2\cos\frac{\pi\kap}{2}$ by $\pi\kern1pt\dkap$. Thus,
\begin{equation}
\dkap \approx \frac{C}{\pi Nt_B (\VF^\A,\VF^\R)}
=\frac{6\alpha_S}{Jk'_\R(-1)\,\Delta(1-\Delta)(1-2\Delta) \tan(\pi\Delta)}\,.
\label{eqn: correction}
\end{equation} 
\end{defn}

\subsection{Maximal chaos in a 1D model}

The quantities involved in the ladder identity may often be regarded as analytic functions of some parameters. Wherever $\kap$ takes on the value of $1$ in the complex domain, the prefactor $C^{-1}$ has a pole. We will discuss some consequences of this fact for SYK-like models in one dimension, where the parameter in question is momentum. A concrete example~\cite{gu2017local} is an array of SYK sites, each containing $N$ Majorana modes. The Hamiltonian includes four-body interactions on each individual site $x$ as well as products of two fermions on site $x$ and two fermions on site $x+1$.

Our goal is to find the connected OTOC of fermionic operators at two different locations:
\begin{equation}
\OTOC_{x,0}(t_1,t_2,t_3,t_4) : = \frac{1}{N^2}\sum_{j,k} \Bigl(\bigr\langle
\chi_{j,x}(t_1)
\chi_{k,0}(t_3)
\chi_{j,x}(t_2)
\chi_{k,0}(t_4)
\bigr\rangle + \langle \chi_{j,x} \chi_{j,x}\rangle \langle \chi_{k,0} \chi_{k,0} \rangle
\Bigr) \,.
\end{equation}
We compute the OTOC through the Fourier transform:
\begin{equation}
\OTOC_{x,0}(t_1,t_2,t_3,t_4) = \int \frac{dp}{2\pi} e^{ipx} \OTOC_p(t_1,t_2,t_3,t_4) \,. \label{eqn: fourier}
\end{equation}
For each momentum eigenmode $\OTOC_p$, the diagrammatics is the same as in the single-site case, and the retarded kernel factorizes:
\begin{equation}
K^\R(p) = s(p)\, K^\R, \quad
s(p) = 1- 2 a (1-\cos p)\,;\qquad
s(p)\approx 1- ap^2\, \text{ for }\, |p|\ll 1\,.
\end{equation}
Loosely speaking, $s(p)$ characterizes the ``band structure'' of the bilocal fields $G_{x}(t_1,t_2)$, and $a$ is a parameter capturing the relative strength of the spatial coupling. (For the nearest neighbor coupling used in \cite{gu2017local}, $a=J_1^2/3J^2 \in (0,1/3)$.) Therefore,
\begin{gather}
k_\R(p,\alpha)=s(p)\,k_\R(\alpha)\approx 1-ap^2+t_B(\alpha+\kap(0)),
\\[3pt]
\kap(p)\approx\kap(0)-t_B^{-1}ap^2,
\label{eq: kapp}
\end{gather}
where $\varkappa(p)$ is obtained by solving the equation $k_\R(p,-\kap(p))=1$. The approximate expressions are based on the assumption that $p$ is small and $\alpha$ is close to $-\kap(0)$. We will see that small $p$ play a dominant role in the butterfly effect if $\kap(0)$ is close to $1$.

Since the ladder identity holds for each momentum eigenmode separately,
\begin{equation}
C(p) = N \cdot 2 \cos \frac{\varkappa (p) \pi}{2} \cdot t_B \cdot (\VF^\A,\VF^\R) \,,
\label{eqn: prefactor}
\end{equation}
the function $\OTOC_p(t_1,t_2,t_3,t_4)\propto C(p)^{-1}$ has a pole at $p$ equal to $p_1=i|p_1|$ such that
\begin{equation}
\kap(p_1)=1\,.
\end{equation}
The dependence of $t_B$ and $(\VF^\A,\VF^\R)$ on $p$ is not important; all we need is that these function are analytic and do not vanish in the domain of interest. In what follows, we take them to be constant. Taking the time dependence of $\OTOC_p(t_1,t_2,t_3,t_4)$ into account, we obtain the following formula:
\begin{equation}
\OTOC_{x,0}(t_1,t_2,t_3,t_4) \approx \frac{1}{N}
\underbrace{\int_{-\infty}^{+\infty} \frac{dp}{2\pi} \,
\frac{e^{\varkappa(p)t+ipx}}{2\cos\frac{\pi\kap(p)}{2}}}_{u(x,t)} \cdot
\frac{\VF^\R(t_{12})\VF^\A(t_{34})}{t_B(\VF^\A,\VF^\R)}\,,\quad\,
t=\frac{t_1+t_2-t_3-t_4}{2}\,.
\label{eq: uxt}
\end{equation}
It is important to remember that the asymptotic form of OTOCs and the related kinetic equation ignore initial correlations as well as any nonlinear effects. Therefore, our calculations are valid if $\OTOC_{x,0}(t_1,t_2,t_3,t_4)$ is much greater than $N^{-1}$ but much less than $1$, that is, in the butterfly wavefront. To determine the butterfly velocity, one may fix an arbitrary value in the indicated range and find at what $x$ and $t$ it is realized. We take it to be on the lower end, where $u(x,t)\sim 1$.

For large $x$ and $t$, the integral in \eqref{eq: uxt} can be estimated by deforming the integration contour in the complex plane so that it passes through a saddle point of the exponent. The saddle point equation $\kap'(p)t+ix=0$ has a purely imaginary solution, $p=i|p|$. If $|p|<|p_1|$, or equivalently, $|x|/t<v_*$, where
\begin{equation}
v_*=i\kap'(p_1),
\end{equation}
the procedure works in a straightforward way. Otherwise the contour has to cross the pole and the integral picks up an additional term. The contributions of the saddle and the pole to $u(x,t)$ are as follows:
\begin{equation}
u_s(x,t) \approx \frac{e^{\kap(p)t+ip|x|}}
{2\cos\frac{\pi\kap(p)}{2}\sqrt{2\pi(-\kap''(p))t}}\,
\text{ where }\, i\kap'(p)=\frac{|x|}{t}\,,\qquad
u_1(x,t) \approx \frac{e^{t+ip_1|x|}}{\pi i\kap'(p_1)}\,.
\label{eq: usu1}
\end{equation}
However, if $|x|/t>v_*$, the first is much smaller than the second.

\begin{figure}[t]
\center
\subfloat[$|p_s|<|p_1|$]{
\begin{tikzpicture}[baseline={(current bounding box.center)}]
\draw[->,>=stealth] (0pt,0pt)--(150pt,0pt) node[right]{\footnotesize $|p|$};
\draw[->,>=stealth] (0pt,0pt) -- (0pt,110pt) node[right]{\footnotesize $\kap(i|p|)$};
\filldraw (0pt,30pt) circle (1pt) node [left]{\footnotesize $\kap(0)$};
\draw (0pt,30pt) .. controls (48pt,30pt) and (91pt,48pt) .. (130pt,90pt);
\draw[dashed, red] (0pt,0pt) -- (120pt,74pt);
\filldraw (87pt,54pt) circle (1pt);
\filldraw (87pt,0pt) circle (1pt) node [below]{\footnotesize $|p_s|$};
\draw[thick, dotted] (87pt,0pt) -- (87pt,54pt);
\filldraw (130pt,90pt) circle (1pt);
\filldraw (130pt,0pt) circle (1pt) node [below]{\footnotesize $|p_1|$};
\filldraw (0pt,90pt) circle (1pt) node [left]{\footnotesize $1$};
\draw[thick, dotted] (0pt,90pt) -- (130pt,90pt);
\draw[thick, dotted] (130pt,0pt) -- (130pt,90pt);
\end{tikzpicture}}
\hspace{30pt}
\subfloat[$|p_s|>|p_1|$]{
\begin{tikzpicture}[baseline={(current bounding box.center)}]
\draw[->,>=stealth] (0pt,0pt)--(150pt,0pt) node[right]{\footnotesize $|p|$};
\draw[->,>=stealth] (0pt,0pt) -- (0pt,110pt) node[right]{\footnotesize $\kap(i|p|)$};
\filldraw (0pt,51pt) circle (1pt) node [left]{\footnotesize $\kap(0)$};
\draw (0pt,51pt) .. controls (48pt,51pt) and (100pt,70pt) .. (150pt,110pt);
\draw[dashed, red] (0pt,0pt) -- (130pt,95pt);
\draw[blue] (0pt,0pt) -- (87pt,71pt);
\filldraw (130pt,95pt) circle (1pt);
\filldraw (130pt,0pt) circle (1pt) node [below]{\footnotesize $|p_s|$};
\draw[thick, dotted] (130pt,0pt) -- (130pt,95pt);
\filldraw (87pt,71pt) circle (1pt);
\filldraw (87pt,0pt) circle (1pt) node [below]{\footnotesize $|p_1|$};
\filldraw (0pt,71pt) circle (1pt) node [left]{\footnotesize $1$};
\draw[thick, dotted] (0pt,71pt) -- (87pt,71pt);
\draw[thick, dotted] (87pt,0pt) -- (87pt,71pt);
\end{tikzpicture}}
\caption{Graphic solution of the equations $\kap(p_1)=1$ and $v_s=i\kap(p_s)/p_s=i\kap'(p_s)$. The velocity $v_s$ is the slope of the red dashed line. In case (a) it is the actual butterfly velocity; in case (b) the butterfly velocity is equal to $v_1=1/|p_1|$, the slope of the blue line.}
\label{fig: imaginary lyapunov}
\end{figure}
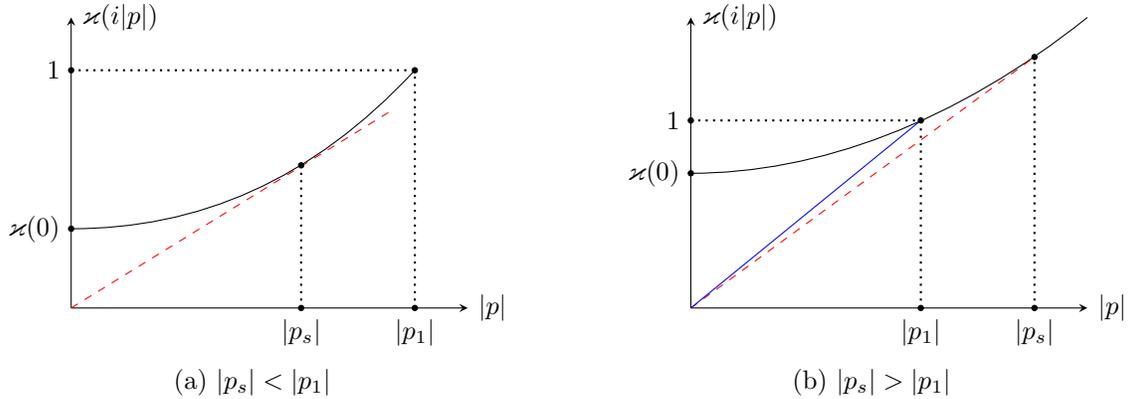

Let us discuss when each situation occurs and determine the butterfly velocity in both cases. The conditions $u_s(x,t)\sim 1$ and $u_1(x,t)\sim 1$ are satisfied, respectively, for $|x|/t\approx v_s$ and $|x|/t\approx v_1$, where
\begin{equation}
v_s=\frac{i\kap(p_s)}{p_s}=i\kap'(p_s)\,,\qquad
v_1=\frac{i}{p_1}\,.
\end{equation}
The solution of these equations is illustrated in figure~\ref{fig: imaginary lyapunov}. If $\kap(0)$ is small, then $|p_s|<|p_1|$, and hence, the pole does not contribute to the OTOC and the butterfly velocity is equal to $v_s$. Conversely, if $\kap(0)$ is close to $1$, then $|p_s|>|p_1|$, the OTOC at large $x$ is dominated by the pole in the prefactor at $p=p_1$, the three velocities satisfy the inequality $v_*<v_s<v_1$, and the butterfly velocity is equal to $v_1$.

From now on, we assume that $\dkap:=1-\kap(0)$ is much less than $1$, while $t_B\sim a\sim 1$. Using the function $\kap(p)$ from \eqref{eq: kapp}, we find that
\begin{equation}
v_* \approx 2\,\sqrt{\frac{a\kern1pt\dkap}{t_B}}\,,\qquad
v_B=v_1 \approx \sqrt{\frac{a}{t_B\kern1pt\dkap}}\,,\qquad
p_1 \approx i\,\sqrt{\frac{t_B\kern1pt\dkap}{a}}\,.
\end{equation}
Note that $p_1\ll 1$, confirming our hypothesis that the relevant values of $p$ are small. Thus, the use of the approximate formula for $\kap(p)$ is justified.

\begin{figure}
[t]
\center
\begin{tikzpicture}[scale=0.8, baseline={(current bounding box.center)}]
\draw[->,>=stealth] (0pt,0pt)--(160pt,0pt) node[right]{\footnotesize $x$};
\draw[->,>=stealth] (0pt,0pt) -- (0pt,160pt) node[above]{\footnotesize $t$};
\draw[thick, blue] (0pt,0pt) -- (140pt,100pt) node[black,  right]{\footnotesize $t=x/v_B$};
\draw[thick, blue] (0pt,65pt).. controls (5pt, 65pt) and (12pt, 69pt) ..(14pt,70pt) -- (140pt,160pt) node[black,  right]{\footnotesize $t=x/v_B+t_{\text{scr}}$};
\draw[thick, olive] (0pt,0pt) -- (26pt,78pt);
\draw[thick, olive, dotted] (26pt,78pt) -- (35pt,105pt) node[black, above]{\footnotesize $t=x/v_*$};
\node at (8pt,50pt){\footnotesize $1$};
\node at (60pt,70pt){\footnotesize $2$};
\node at (60pt,145pt){\footnotesize $3$};
\node at (100pt,20pt){\footnotesize $4$};
\end{tikzpicture}
\caption{
Four regions with different OTOC behavior.
}
\label{fig: four regions}
\end{figure}
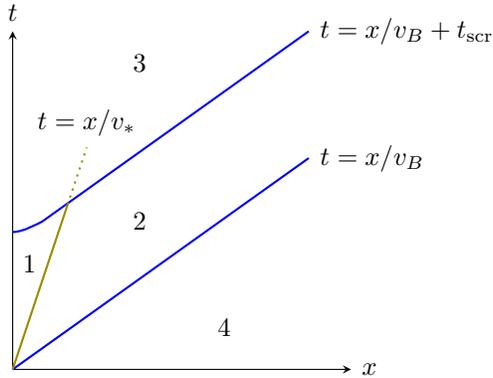

A complete solution for all $x$ and $t$ would require including nonlinearity in the kinetic equation. However, the following seems to be a consistent qualitative picture. Let us define the scrambling time as $t_{\text{scr}}=\ln(\dkap\kern1pt N)$. There are four regions for the butterfly effect as shown in figure~\ref{fig: four regions}:

\begin{enumerate}
\item $\frac{x}{v_*} < t< \frac{x}{v_B}+t_{\text{scr}}$. In this region, the saddle contribution to the OTOC exceeds that of the pole; hence
\begin{equation}
\OTOC \sim \frac{u_s(x,t)}{N}  \approx
\frac{1}{\pi\kern1pt\dkap\kern1pt N}\,
\frac{e^{\kap(0)t-x^2/(4D_st)}}{\sqrt{4\pi D_st}},\,\qquad
D_s=\frac{a}{t_B}\,.
\end{equation}
The Lyapunov exponent $\kap(0)$ is the same as for the usual SYK model.

\item $\frac{|x|}{v_B}<t<\frac{|x|}{v_B}+t_{\text{scr}}$ and $t<\frac{|x|}{v_*}$. This is the most interesting region. It is dominated by the contribution from the pole, which grows with the maximal Lyapunov exponent:
\begin{equation}
\OTOC \sim \frac{u_1(x,t)}{N}  \approx
\frac{1}{\pi v_*N}\, e^{t- |p_1| |x|} \,.
\end{equation}

\item $t>\frac{|x|}{v_B}+t_{\text{scr}}$. In this region the $\OTOC$ has saturated.

\item $t< \frac{|x|}{v_B}$. In this region the $\OTOC$ is smaller than $1/N$ and the butterfly effect is negligible.
\end{enumerate}

We remark that the exact maximal Lyapunov exponent here arises from the zero of the decoherence factor $2\cos\frac{\varkappa\pi}{2}$, which is related to the pole of the prefactor by the ladder identity we derived. This argument is stronger than the perturbative calculation done in~\cite{gu2017local}, where the authors note a cancellation of corrections to the Lyapunov exponent at order $1/J$. As we have demonstrated, the Lyapunov exponent in the butterfly wavefront is exactly $1$, provided the coupling strength $J$ is above a certain threshold. It would be interesting to see if another perturbative result in~\cite{gu2017local} is actually exact, namely, that the butterfly velocity is related to the heat diffusion coefficient as $v_B^2=D$. This relation has previously been established for certain holographic theories~\cite{Blake2016}.

Finally, we comment on the commutator OTOC. Naively, it could be expected to propagate with velocity $v_s<v_B$ because the corresponding prefactor, equal to $\frac{2\cos(\varkappa\pi/2)}{C}$~\cite{KS17-soft}, has no pole. However, commutator OTOCs also contain terms proportional to $\bigl(C^{-1}e^{\kap t}\bigr)^2$ due to nonlinearity. We conjecture that under the previous assumptions, there is only one butterfly velocity $v_B=v_1$, though the commutator OTOC has a different wavefront profile.

\section{Commutator OTOCs and stringy states}\label{sec: stringy}

Out-of-time-order correlators that involve two commutators (for bosons) or two anticommutators (for fermions) capture some interesting physics. In the black hole setting~\cite{ShSt14}, the connected OTOC represents the gravitational scattering amplitude between infalling and emerging particles, whereas the commutator OTOC corresponds to the full scattering probability. This interpretation (but not the actual calculation) can be extended to other systems by a suitable definition of in- and out-states~\cite{KS17-soft}, see  \eqref{eq: OTOC-scattering}, \eqref{eq: OTOCR-scattering} below. In the Born approximation, i.e.\ at early times, the probability of elastic scattering is proportional to $\bigl(C^{-1}e^{\kap t}\bigr)^2$. Ignoring such nonlinear terms, the commutator OTOC is determined by inelastic scattering. We will express the commutator OTOC for the SYK model as the inner product between two vectors in a suitable Hilbert space. In a loose sense, such vectors represent the products of inelastic collisions. We call them ``stringy states'' because they are generated from some vacuum by strings of bosonic operators $\calO_{jk}$. While our construction is an effective model, the operators in question may be identified with the microscopic observables
\begin{equation}
\calO_{jk}=i^{q/2-1}\sum_{l_1<\cdots<l_{q-2}}
J_{jkl_1\ldots l_{q-2}}\chi_{l_1}\cdots\chi_{l_{q-2}}.
\end{equation}
Such operators were proposed as an interpretation of a certain term in the OPE of $\chi_j(\tau_1)\chi_k(\tau_2)$ and a conjecture was made that is analogous to our result in the conformal limit~\cite{Douglas_note}. A similar OPE calculation was done in~\cite{Josephine_note}.

\begin{defn}[Retarded OTOC.] This is a particular variant of commutator OTOC:
\begin{equation}
\OTOC^\R(t_1,t_2,t_3,t_4) := \theta(t_{13})\theta(t_{24})\cdot
\frac{1}{N^2}\sum_{j,k} \bigl\langle
\{\chi_{j}(it_1+\pi),\chi_{k}(it_3+\pi)\}
\{\chi_{j}(it_2),\chi_{k}(it_4)\}
\bigr\rangle\,.
\label{eqn: retarded OTOC}
\end{equation}
We will show, following~\cite{MSW17}, that it equals $N^{-1}$ times the sum of all ladders with the rails made of retarded Green functions and the rungs of Wightman functions.

Before proving this statement, let us remark that it provides an alternative route to the ladder identity. As a first step, expand both anticommutators in \eqref{eqn: retarded OTOC} and consider the limit of large $t:=\frac{t_1+t_2-t_3-t_4}{2}$. There are two terms with alternating times, $1324$ and $3142$, and two other terms, $3124$ and $1342$. The former obey the asymptotic formula \eqref{OTOC1} while the latter obey the approximate clustering property. In the resulting expression, all disconnected correlators cancel, and we get
\begin{equation}
\OTOC^\R(t_1,t_2,t_3,t_4) \approx
r\, e^{\varkappa (t_1+t_2-t_3-t_4)/2} \VF^\R(t_{12}) \VF^\A (t_{34})\,,\qquad
r=\frac{2\cos\frac{\kap\pi}{2}}{C}\,.
\end{equation}
(The calculation involved is similar to that of $\OTOC_A+\OTOC_{A'}$ in \eqref{eqn: OTOCA}, \eqref{eqn: OTOCA'}.) According to the hypothesis, the retarded OTOC is simply a sum of ladders, and the cutting argument used to derive the ladder identity yields this formula:
\begin{equation}
Nrk'_\R(-1)\bigl(\VF^\A,\VF^\R)=1.
\end{equation}

\begin{figure}[t]
\center
\begin{tikzpicture}[scale=0.8]
\draw[->,>=stealth] (-80pt,0pt) -- (80pt,0pt) node[right] {\scriptsize $\Re e^{i\theta}$};
\draw[->,>=stealth] (0pt,-80pt) -- (0pt,80pt) node[right]{\scriptsize $\Im e^{i\theta}$};

\draw[thick,blue] (7pt,3pt)--(60pt,3pt);
\draw[thick,blue] (-7pt,3pt)--(-60pt,3pt);
\draw[thick,blue] (7pt,-3pt)--(60pt,-3pt);
\draw[thick,blue] (-7pt,-3pt)--(-60pt,-3pt);

\draw[blue,thick] (7pt,3pt) arc (90:270:3pt);
\draw[blue,thick] (-7pt,3pt) arc (90:-90:3pt);
\draw[blue,thick,dashed] (60pt,3pt) arc (4:176:60pt);
\draw[blue,thick,dashed] (60pt,-3pt) arc (-4:-176:60pt);

\filldraw[fill=black] (15pt,-3pt) circle (1pt) node[below]{\scriptsize $2$};
\filldraw[fill=black] (-15pt,-3pt) circle (1pt) node[below]{\scriptsize $1$};

\filldraw[fill=black] (50pt,-3pt) circle (1pt) node[below]{\scriptsize $4$};
\filldraw[fill=white] (50pt,3pt) circle (1pt) node[above]{\scriptsize $4'$};

\filldraw[fill=black] (-50pt,3pt) circle (1pt) node[above]{\scriptsize $3$};
\filldraw[fill=white] (-50pt,-3pt) circle (1pt) node[below]{\scriptsize $3'$};
\end{tikzpicture}
\caption{The contour for the retarded OTOC. The dashed circle is assumed to be infinitely big so that the folds cover the intervals $(-\infty,0)$ and $(0,+\infty)$.}
\label{fig: retarded OTOC}
\end{figure}
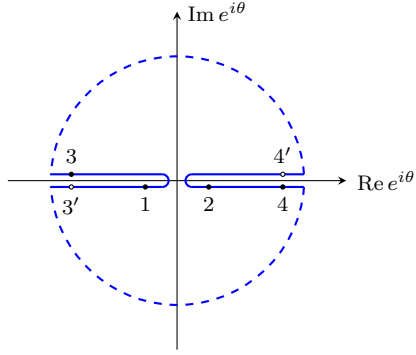

To obtain a diagrammatic expression for the retarded OTOC, we represent it as a linear combination of four correlators on the contour shown in figure~\ref{fig: retarded OTOC}:
\begin{equation}
\begin{aligned}
N\OTOC^\R(t_1,t_2,t_3,t_4)=
&-\calF(\theta_1,\theta_2,\theta_3,\theta_4)
+\calF(\theta_1,\theta_2,\theta_3',\theta_4)\\
&+\calF(\theta_1,\theta_2,\theta_3,\theta_4')
-\calF(\theta_1,\theta_2,\theta_3',\theta_4')\,,\\
\theta_1 \equiv it_1+\pi, \
\theta_2 \equiv it_2,\ &
\theta_3 \equiv it_3+\pi - \epsilon , \
 \theta_3' \equiv it_3+\pi+\epsilon, \
\theta_4 \equiv it_4 -\epsilon , \
\theta_4' \equiv it_4+\epsilon \,.
\end{aligned}
\label{eq: 4corr}
\end{equation}
Each of the four correlators in \eqref{eq: 4corr} satisfies the Bethe-Salpeter equation \eqref{eqn: kernel equation} with the same kernel $K$ but different free term $\calF_0$. Therefore, their linear combination, considered as a function of $\theta_1,\theta_2$, satisfies a similar equation. Since this function does not distinguish between forward and backward contour branches (i.e.\ the upper and lower parts of the folds), the kernel is reduced to the retarded one. Meanwhile, the free terms add up to $-G^\R(t_1,t_3)G^\R(t_2,t_4)$. Thus,
\begin{equation}
\begin{aligned}
N\OTOC^\R(t_1,t_2,t_3,t_4) =& -G^\R(t_1,t_3)G^\R(t_2,t_4)\\
& +\int_{\RR} dt_5\,dt_6\, K^\R (t_1,t_2,t_5,t_6)\kern1pt
N\OTOC^\R(t_5,t_6,t_3,t_4)\,.
\end{aligned}
\end{equation}
The solution of this equation is the sum of retarded ladders.
\end{defn}

\begin{defn}[In- and out-states.] Let $\calH$ be the Hilbert space of some quantum system and $\calH^*$ the dual space. We denote basis vectors of $\calH$ as $|n\rangle$ and those of $\calH^*$ as $|\bar{n}\rangle$. More generally, for each $|\psi\rangle=\sum_{n}c_n|n\rangle\in\calH$ there is a corresponding dual vector $|\bar{\psi}\rangle=\sum_{n}c_n^*|\bar{n}\rangle$. Using this notation, it is easy to convert operators acting in $\calH$ to vectors in $\calH\otimes\calH^*$:
\begin{equation}
A=\sum_{n,m}A_{nm}|n\rangle\langle m|\quad\mapsto\quad
|A\rangle:=\sum_{n,m}A_{nm}|n\rangle\otimes|\bar{m}\rangle\,.
\end{equation}
In particular, if $\rho=Z^{-1}e^{-2\pi H}$ is the thermal state corresponding to some Hamiltonian $H$, then $|\rho^{1/2}\rangle$ is the thermofield double state. We are interested in states that can be obtained from the thermofield double by the action of some simple operators. There are two natural way to apply an operator $X$, which give $(X\otimes I)|\rho^{1/2}\rangle=|X\rho^{1/2}\rangle$ and $(I\otimes X^T)|\rho^{1/2}\rangle=|\rho^{1/2}X\rangle$. For consistency with our previous notation, we will use a symmetric definition, $|\rho^{1/4}X\rho^{1/4}\rangle$. Note that
\begin{equation}
\bigl\langle\rho^{1/4}Y\rho^{1/4}\big|\rho^{1/4}X\rho^{1/4}\bigr\rangle
=\Tr\bigl(\rho^{1/2}Y^\dag\rho^{1/2}X\bigr) =
\bigl\langle Y^\dag(\pi)X(0)\bigr\rangle\,,
\end{equation}
where the last bracket denotes the thermal average.

From now on, we consider a system with a large parameter $N$, such as the SYK model. Let $\kap^{-1}\ll t_{+}-t_{-}\lesssim t_{\text{scr}}$, and let us define the following states:\footnote{Alternatively, one could consider the states $|\chi_j(it_+)\rho^{1/2}\chi_k(it_-)\rangle$ and $-|\chi_k(it_-)\rho^{1/2}\chi_j(it_+)\rangle$, each describing a pair of particles produced at opposite boundaries of a two-sided black hole.
However, such states are less convenient because in the absence of interaction, they are not equal to each other.}
\begin{equation}
\bigl|\psi^{\text{out}}_{jk}(t_{+},t_{-})\bigr\rangle =
\bigl|\rho^{1/4}\chi_{j}(it_+)\chi_{k}(it_-)\rho^{1/4}\bigr\rangle\,,\quad\,
\bigl|\psi^{\text{in}}_{jk}(t_{+},t_{-})\bigr\rangle =
-\bigl|\rho^{1/4}\chi_{k}(it_-)\chi_{j}(it_+)\rho^{1/4}\bigr\rangle\,.
\end{equation}
The inner products of states of the same type are well approximated using a naive model that replaces the thermofield double with a suitable vacuum of non-interacting fermions. For example,
\begin{equation}
\bigl\langle\psi^{\text{out}}_{jk}(t_1,t_3)\big|
\psi^{\text{out}}_{jk}(t_2,t_4)\bigr\rangle
\approx G^\W(t_1,t_2)G^\W(t_3,t_4)
\label{eq: out-out product}
\end{equation}
up to $1/N$ corrections that do not grow with time. However, this model is not very accurate at reproducing the inner product between an out-state and an in-state:
\begin{equation}
\begin{aligned}
\bigl\langle\psi^{\text{out}}_{jk}(t_1,t_3)\big|
\psi^{\text{in}}_{jk}(t_2,t_4)\bigr\rangle
&= -\Tr\bigl(\chi_j(it_1)\rho^{1/2}\chi_k(it_4)\chi_j(it_2)
\rho^{1/2}\chi_k(it_3)\bigr)\\
&= G^\W(t_1,t_2)G^\W(t_3,t_4)
-\OTOC\biggl(t_1,t_2,\,t_4+i\frac{\pi}{2},\,t_3+i\frac{\pi}{2}\biggr)
\,.
\end{aligned}
\label{eq: out-in product}
\end{equation}
(This equation is approximate for each particular $j$ and $k$ and exact if averaging is performed.) The second term may be interpreted as a scattering amplitude in the naive model~\cite{KS17-soft}.

This formula defines both elastic and and inelastic scattering:
\begin{equation}
\bigl|\psi^{\text{in}}_{jk}(t_+,t_-)\bigr\rangle 
-\bigl|\psi^{\text{out}}_{jk}(t_+,t_-)\bigr\rangle
=\bigl|\psi^{\text{el}}_{jk}(t_+,t_-)\bigr\rangle
+\bigl|\psi^{\text{inel}}_{jk}(t_+,t_-)\bigr\rangle\,,
\label{eq: el-inel decomposition}
\end{equation}
where $|\psi^{\text{el}}_{jk}(t_+,t_-)\rangle$ is a linear combination of out-states and $|\psi^{\text{inel}}_{jk}(t_+,t_-)\rangle$ is orthogonal to all such states. Left-multiplying \eqref{eq: el-inel decomposition} by $\langle\psi^{\text{out}}|$ and by $\langle\psi^{\text{in}}|-\langle\psi^{\text{out}}|$ with suitable parameters, we get:
\begin{align}
-\OTOC\biggl(t_1,t_2,\,t_4+i\frac{\pi}{2},\,t_3+i\frac{\pi}{2}\biggr)
&\approx\bigl\langle\psi^{\text{out}}_{jk}(t_1,t_3)\big|
\psi^{\text{el}}_{jk}(t_2,t_4)\bigr\rangle\,,
\label{eq: OTOC-scattering}\\[3pt]
\OTOC^\R(t_1,t_2,t_3,t_4)
&=\bigl\langle\psi^{\text{el}}_{jk}(t_1,t_3)\big|
\psi^{\text{el}}_{jk}(t_2,t_4)\bigr\rangle
+\bigl\langle\psi^{\text{inel}}_{jk}(t_1,t_3)\big|
\psi^{\text{inel}}_{jk}(t_2,t_4)\bigr\rangle\,.
\label{eq: OTOCR-scattering}
\end{align} 
Focusing on the early times, when both OTOCs are small due to the $N^{-1}$ factor, we conclude that the norms of $|\psi^{\text{el}}\rangle$ and $|\psi^{\text{inel}}\rangle$ are proportional to $N^{-1}$ and $N^{-1/2}$, respectively. Thus in this case, the retarded OTOC is dominated by the inelastic term.
\end{defn}
\begin{figure}[t]
\center
\begin{tikzpicture}[scale=0.8]
\draw (-100pt,-40pt) -- (50pt,-40pt);
\filldraw  (-100pt,-40pt) circle (1pt) node[below] {\footnotesize $t_+$};
\filldraw (-40pt,-40pt) circle (1pt) node[below] {\footnotesize $t_1$};
\filldraw (20pt,-40pt) circle (1pt) node[below] {\footnotesize $t_2$};
\draw (-40pt,-40pt) ..controls (-50pt,-20pt) and (-50pt,20pt) .. (-50pt,20pt) ;
\draw (-40pt,-40pt) ..controls (-30pt,-20pt) and (-30pt,20pt) .. (-30pt,20pt) ;
\draw (20pt,-40pt) ..controls (30pt,-20pt) and (30pt,20pt) .. (30pt,20pt) ;
\draw (20pt,-40pt) ..controls (10pt,-20pt) and (10pt,20pt) .. (10pt,20pt) ;
\node at (-70pt,-40pt) [above] {\scriptsize $j$};
\node at (-10pt,-40pt) [above] {\scriptsize $s_1$};
\node at (50pt,-40pt) [above] {\scriptsize $s_2$};
\draw [thick, loosely dotted] (53.5pt,-40pt) -- (100pt,-40pt);
\draw (100pt,-40pt) -- (250pt,-40pt);
\node at (100pt,-40pt) [above] {\scriptsize $s_{n-2}$};
\node at (160pt,-40pt) [above] {\scriptsize $s_{n-1}$};
\node at (220pt,-40pt) [above] {\scriptsize $k$};
\filldraw  (250pt,-40pt) circle (1pt) node[below] {\footnotesize $t_-$};
\filldraw (190pt,-40pt) circle (1pt) node[below] {\footnotesize $t_{n}$};
\filldraw (130pt,-40pt) circle (1pt) node[below] {\footnotesize $t_{n-1}$};
\draw (190pt,-40pt) ..controls (180pt,-20pt) and (180pt,20pt) .. (180pt,20pt) ;
\draw (190pt,-40pt) ..controls (200pt,-20pt) and (200pt,20pt) .. (200pt,20pt) ;
\draw (130pt,-40pt) ..controls (120pt,-20pt) and (120pt,20pt) .. (120pt,20pt) ;
\draw (130pt,-40pt) ..controls (140pt,-20pt) and (140pt,20pt) .. (140pt,20pt) ;
\end{tikzpicture}
\caption{Stringy state corresponding to a half of a retarded ladder.}
\label{fig: stringy}
\end{figure}
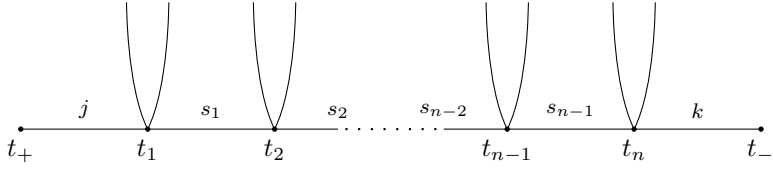

\begin{defn}[Stringy states.] Now we give an effective model for the inelastic scattering. Let $\calO_{jk}=-\calO_{kj}$ be non-interacting bosons with the thermal correlator
\begin{equation}
\bigl\langle\T\calO_{j'k'}(\tau')\calO_{jk}(\tau)\bigr\rangle
=\frac{J^2(q-1)}{N-1}G(\tau',\tau)^{q-2}
\bigl(\delta_{j'j}\delta_{k'k}-\delta_{j'k}\delta_{k'j}\bigr)\,.
\end{equation}
\end{defn}
We denote the corresponding thermal state by $\rho_{\text{eff}}$, call $\bigl|\rho_{\text{eff}}^{1/2}\bigr\rangle$ ``vacuum'', and consider excited states generated by strings of operators:
\begin{equation}
\bigl|\psi_{s_0\ldots s_n}(t_1,\ldots,t_n)\bigr\rangle := 
\Bigl|\rho_{\text{eff}}^{1/4}\,
\calO_{s_0s_1}(it_1)\cdots\calO_{s_{n-1}s_n}(it_n)\,
\rho_{\text{eff}}^{1/4}\Bigr\rangle\,.
\end{equation}
The inner product of two such vectors, summed over indices, is a product of Wightman functions:
\begin{equation}
\begin{aligned}
&\frac{1}{N^2}\sum_{j,k}\sum_{s_1',\ldots,s_{n-1}'}\sum_{s_1,\ldots,s_{n-1}}
\bigl\langle\psi_{js_1'\ldots s_{n-1}'k}(t_1',\ldots,t_n')
\big|\psi_{js_1\ldots s_{n-1}k}(t_1,\ldots,t_n)\bigr\rangle\\
&\qquad=\frac{1}{N}\Bigl(J^2(q-1)G^\W(t_1',t_1)^{q-2}\Bigr)
\cdots\Bigl(J^2(q-1)G^\W(t_n',t_n)^{q-2}\Bigr)\,.
\end{aligned}
\end{equation}
Finally, we postulate the following equation, illustrated in figure~\ref{fig: stringy}:
\begin{equation}
\bigl|\psi^{\text{inel}}_{jk}(t_+,t_-)\bigr\rangle
=\sum_{n=0}^{\infty}\int dt_1\cdots dt_n \!\sum_{s_1,\ldots,s_{n-1}} \underbrace{G^\R(t_+,t_1)\cdots G^\R(t_n,t_-)
\bigl|\psi_{js_{1}\ldots s_{n-1}k}(t_1,\ldots,t_n)\bigr\rangle}
_{\text{half of a retarded ladder}}.
\label{eq: stringy}
\end{equation}
The degenerate $n=0$ term should be understood as $G^\R(t_+,t_-)\,\delta_{jk}\bigl|\rho_{\text{eff}}^{1/2}\bigr\rangle$. The inner product of such states is a sum of retarded ladders. More exactly,
\begin{equation}
\frac{1}{N^2}\sum_{j,k}\bigl\langle\psi^{\text{inel}}_{jk}(t_+',t_-')
\big|\psi^{\text{inel}}_{jk}(t_+,t_-)\bigr\rangle =
\OTOC^\R(t_+',t_+,t_-',t_-)\,.
\end{equation}
In fact, an approximate equality holds even if we do not average over $j$ and $k$. Thus, the states \eqref{eq: stringy} correctly reproduce the inner products between actual inelastic scattering states, defined in \eqref{eq: el-inel decomposition}.

\section{Summary}

Our results constitute a step toward an effective theory of OTOCs for systems with dominant ladder diagrams. We have gained some insight into ``stringy effects'', which are characterized by the retarded OTOC, equal to the sum of retarded ladder. Their effective strength has been measured as $t_B^{-1}$, the average number of rungs per unit time. The ladder identity, $N\frac{2\cos(\varkappa\pi/2)}{C}t_{B}(\VF^A,\VF^R)=1$, provides a link between the prefactor $C$ of the connected OTOC and the Lyapunov exponent $\kap$. For the SYK model, it extends the theory of maximal chaos, where $C$ is obtained from the Schwarzian action, to near-maximal chaos. Specifically, it gives the correction to the Lyapunov exponent because $t_B$, $\VF^A$, and $\VF^R$ are well-defined in the conformal limit.

As a somewhat mysterious corollary of the ladder identity, maximal chaos occurs in the butterfly wavefront of an SYK-like one-dimensional model, provided the parameter $J$ is above threshold.

\section*{Acknowledgments}

We thank 
David Huse,
Juan Maldacena, 
Xiao-Liang Qi,
Subir Sachdev,
Steve Shenker,
Douglas Stanford,
Josephine Suh
and 
Ashvin Vishwanath
for useful discussions.
Y.G.\ is supported by the Gordon and Betty Moore Foundation EPiQS Initiative through Grant (GBMF-4306). A.K.\ is supported by the Simons Foundation under grant~376205 and through the ``It from Qubit'' program, as well as by the Institute of Quantum Information and Matter, a NSF Frontier center funded in part by the Gordon and Betty Moore Foundation. This work was performed in part at Aspen Center for Physics, which is supported by National Science Foundation grant PHY-1607611, and at KITP, supported by the NSF grant PHY-1748958.

\bibliography{ref.bib}

\end{document}